\documentclass[5p]{elsarticle}
\journal{Physics Letters B}

\usepackage{graphicx}
\usepackage{amssymb}
\usepackage{epstopdf}
\usepackage{verbatim}
\usepackage{amsthm,  mathrsfs, enumerate}
\usepackage{amsmath}

\newcommand{\GeV}{\ensuremath{\,{\rm GeV}}}

\newcommand{\beq}{\begin{equation}}
\newcommand{\eeq}{\end{equation}}

\newcommand{\req}[1]{Eq.\,(\ref{#1})}
\newcommand{\order}[1]{ \mathcal{O} \left( #1 \right) }

\begin{document}

\begin{frontmatter}

\title{Possibility of Electroweak Phase Transition at Low Temperature}

\author{Jeremiah Birrell$^1$}
\author{and Johann Rafelski$^2$}
\address{$^1$Program in Applied Mathematics and $^2$Department of Physics\\
 The University of Arizona, Tucson, Arizona, 85721, USA}

\date{May 4, 2012}

\begin{abstract}
We study models of strong first order  `low' temperature  electroweak phase transition. To achieve this we propose a class of Higgs effective potential models which preserve  the known features of the present day massive phase. However, the properties of the symmetry restored massless phase are modified in a way that for a large parameter domain we find a strong first order transition occurring at a temperature hundreds of times lower than previously considered possible.  
\end{abstract}


\end{frontmatter}

{\bf Introduction:}
In the early universe at sufficiently high temperature $T$ the electroweak (EW) symmetry is restored and as result elementary  particles existed in a massless phase\cite{Linde:2005ht,Sher:1988mj,Shaposhnikov:1993tv,Djouadi:2005gi}. There is general agreement that  for the parameter set of the Standard Model (SM) quantum corrections wash out the weak 1st order phase transition\cite{Kajantie:1996mn,Rummukainen:1998as,Csikor:1998eu,Fodor:1998ji,Tsypin:1998gv}.   However, seeing the Higgs mechanism as an effective theory opens the opportunity to modify the Higgs effective potential.  Such a modification can have significant impact on the physics of the electroweak phase transition, for example making it strong\cite{Grojean:2004xa}. We demonstrate by example that a strong transition is possible at a temperature which can be a small  fraction of the usual EW energy scale.

Considering the SM Higgs sector as an effective theory, the theoretical constraints on the potential shape from the condition of  renormalizibility should  be applied solely in the neighborhood of the different vacuum states where perturbation theory applies. This removes the requirement that the Higgs potential $W(h)$ contain only quadratic and quartic terms in $h$ and motivates our study. We model here the shape of the Higgs effective potential $W(h)$ away from the vacuum expectation value of $\langle h\rangle =v_0$ where it has not been probed experimentally. 

The established characteristics of the SM Higgs potential $V(h)$ in the unitary gauge are that it possesses a minimum at $v_0\equiv 246/\sqrt{2}$ GeV with $V(v_0)=0$  and a mass constrained near $m_h\simeq 125 \GeV$ according to the latest LHC results\cite{Ellis:2012rx}.  This determines the second derivative at $h=v_0$.  In the unitary gauge, the potential is typically taken to be symmetric and quadratic in $h^2$,
\begin{equation}\label{standard_higgs}
V(h)=\frac{\lambda_h}{4}(h^2-v_0^2)^2, \hspace{2mm} \lambda_h=\frac{m_h^2}{2v_0^2}
\end{equation}
which as noted produces a renormalizable theory, but this is far from the unique choice with the required characteristics.

\vskip 0.2cm 
{\bf An effective Higgs potential model due to unknown physics:}
Viewing the electroweak (EW) sector of SM as an effective theory, we have no reason to constrain the behavior  of the effective potential $W$  away  from $v_0$ by the form \req{standard_higgs}. Our rationale for employing a different Higgs effective potential is the same as offered in Ref.\,\cite{Grojean:2004xa}, i.e. a consequence of the elimination/integration out  of other dynamical and/or auxiliary fields reflecting a possible bound state structure of Higgs. We refrain from entering into detailed discussion of how an effective potential arises  in favor of an exploration of consequences of effective potential variants.

We consider the following family of potentials
\begin{equation}\label{modified_higgs}
W(h)=f(V(h)), \qquad f(x)=\frac{x}{(1+B x/V(0))^k}.
\end{equation}
This form \req{modified_higgs} preserves the known vacuum expectation value and mass of the Higgs, but modifies the potential near $h=0$ and for large values of $h$.  Note that this is just one possible modification. We could take $f$ in \req{modified_higgs} to be any smooth function with $f(0)=0$ and $f^{'}(0)=1$ and still maintain the desired properties of the Higgs.

By choosing $f$ to be analytic in a neighborhood of the real axis, our expansion near the vacuum state contains only even powers in $h^2-v_0^2$, 
\begin{equation}
W(h)=V(h)-\frac{\lambda Bk}{4v_0^4}(h^2-v_0^2)^4+\order{(h^2-v_0^2)^6}.
\end{equation}
differing in this detail from Ref.\,\cite{Grojean:2004xa}  which obtains their results by adding $(h^2-v_0^2)^3$  to the SM Higgs  potential.   The new physics which we find is primarily due to the modification of the potential in the neighborhood of the massless phase, not its perturbative properties in the neighborhood of $v_0$. Since our potential is not a polynomial, most of our findings are novel or/and  differ from\cite{Grojean:2004xa}.

\begin{figure}[tb]
\includegraphics[height=8.8cm]{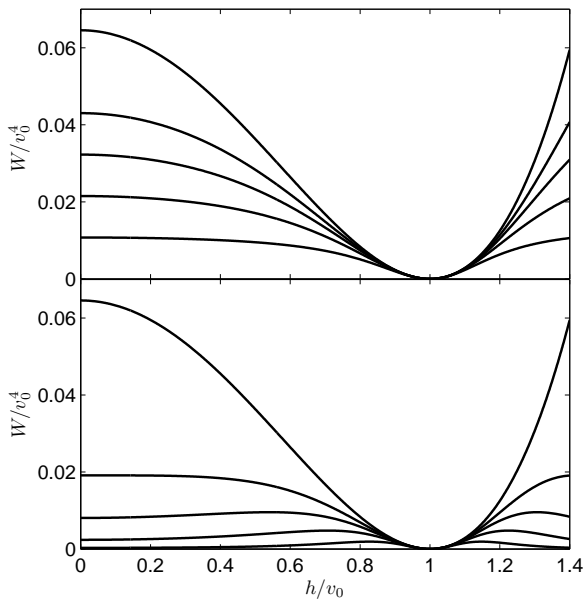}
\caption{Higgs potential for $k=1$ (top) and $k=3$ (bottom) and for $B=0,1/2,1,2,5$ (top curve to bottom curve).\label{fig:W_potential}}
\end{figure}

As shown in Fig.~\ref{fig:W_potential}, the parameter $B$  in \req{modified_higgs} controls the height of the potential at $h=0$,
\begin{equation}\label{Bmeaning}
B=(V(0)/W(0))^{1/k}-1.
\end{equation}
We restict to $B>0$ in order to avoind singularities in the potential.  In this case, $W(0)<V(0)$ and so the height of the potential at the origin is lower compared to \req{standard_higgs}.  This allows for effects at lower energy scales than predicted by the standard potential.  For $k=1$, $W\underset{h\rightarrow \infty}{\longrightarrow} V(0)/B$.  This marks the boundary between stability and instability.  When $k>1$, the height of the potential at $h=0$ varies with $B$, but the qualitative behavior of the potential for $B>0$ is the same as for $B=0$.  In particular, $h=\pm v_0$ are degenerate stable equilibrium points.  When $k> 1$, the collection of equilibrium points $h=\pm v_0$ ceases to be a globally stable set, since $W\underset{h\rightarrow \infty}{\longrightarrow}0$. 

More significantly, when
\begin{equation}\label{Tc1_curve}
(k-1)B>1
\end{equation}
the critical point at $h=0$ becomes a local minimum rather than a local maximum, as can be seen in the bottom panel of  Fig.~\ref{fig:W_potential}.  This state is the massless phase of the theory.  The only massive particle in the symmetry restored massless phase is the Higgs. In our model we find  
\begin{equation}
\tilde{m}_h^2=m_h^2\frac{(k-1)B-1}{2(1+B)^{k+1}}
\end{equation}
where $m_h$ is the Higgs mass in the symmetry broken massive phase.  Also note that the boundary case $k=1+1/B$ the `massless' phase corresponds to an $h^4$-model.

\vskip 0.2cm 
{\bf Finite temperature modification of the effective Higgs potential due to SM particles:}
The minimal coupling of all SM particles to the Higgs contributes a strong effective potential modification. This has been studied in depth for the SM Higgs potential $V(h)$ and it is found that for $T=\order{v_0}$  there is a transition or transformation of the  vacuum ground state into the massless phase. Without quantum corrections a weak 1st order transition  was found\cite{Linde:2005ht,Shaposhnikov:1993tv}.  However,  numerical lattice simulation  showed a more complex structure. For Higgs mass in the physical domain there is no phase transition, but a crossover between the massive and symmetry restored phases\cite{Kajantie:1996mn,Rummukainen:1998as,Csikor:1998eu,Fodor:1998ji,Tsypin:1998gv}. Since the 1st order phase transition we find in our model will be strong, we believe that the results we present in the lowest order of Higgs-matter coupling will remain valid upon inclusion of  quantum-fluctuation effects.
 
If we model each elementary particle species as a Bose or Fermi gas, then each species contributes a free energy density term $F=(-T\ln Z)/V$, and we explore the combined effective potential  $U$
\beq\label{TdepW}
U(h,T)=W(h)-W(0)+\sum_j [F_{\pm,j}(h,T)-F_{\pm,j}(0,T)].
\eeq
For the purpose of characterizing a phase transition between massive and massless phases, only differences in the potential are important and thus we removed from $U$ the solely $T$-dependent shift, normalizing the value at the origin to $U(0,T)=0$. The free energy density terms are\cite{Sher:1988mj}
\begin{align}\label{F_no_shift}      
F_{\pm}(h,T)&=\mp\frac{g_{s} T^4}{2\pi^2}\int_0^\infty\!\!\!\! \ln\left(1\pm e^{-E/T}\right)z^2 dz,\\[0.2cm]
E/T&=\sqrt{z^2+g^2h^2/T^2},\qquad z=p/T.\notag 
\end{align}
Here $g_s$ is the degeneracy, $g$ the strength of particle-Higgs coupling, `plus' is for fermions, and `minus' for bosons.  The difference in free energy between the massive and massless  phases is 
\begin{equation}\label{FreeT}
F_{\pm}(h,T)-F_{\pm}(0,T)=\mp\frac{g_{s} T^4}{2\pi^2}\int_0^\infty \!\!\!\!\ln\left(\frac{1\pm e^{-E/T}}{1\pm e^{-z}}\right)z^2 dz.
\end{equation}

\begin{figure}
\includegraphics[height=9cm]{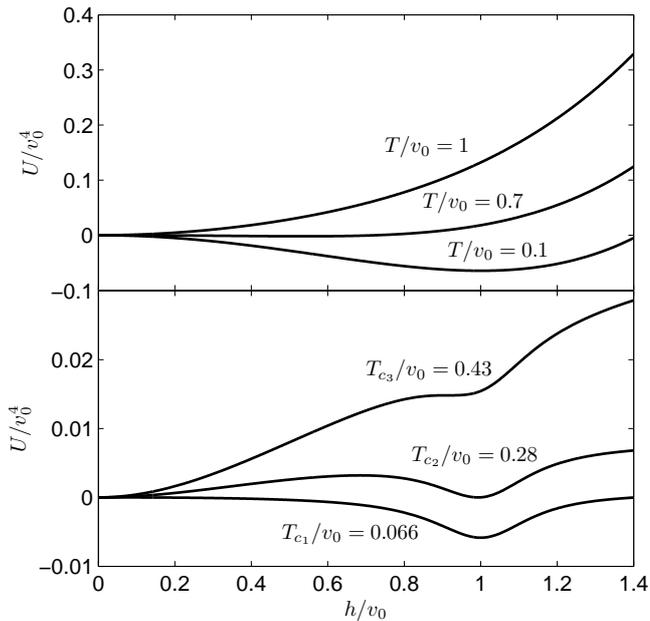}
\caption{Higgs Effective Potential for  $B=0$, $k=1$ (top) and $B=10$, $k=1$ (bottom) for Higgs-top quark system at finite temperature.\label{fig:eff_pot}}
\end{figure}

It is very instructive to first note that the unit coupling of the top quark to Higgs  $v_0\simeq \sqrt{2}m_t$ can alone have a very significant impact on the  effective potential, a point which was not part of the earlier investigations of the phase structure of EW theory.  In Fig.~\ref{fig:eff_pot} we show the effective potential solely derived from inclusion of a top quark gas, in the upper   frame for the case of SM Higgs potential, and in bottom frame for the case of $k=1, B=10$ which indicates at $T=0.43 v_0$ a strong 1st order phase transition in which $h\simeq v_0$ switches to massless vacuum $h=0$. Thus the strongly coupled top-Higgs system  produces the desired phase property derived earlier from Higgs-W-Z dynamics.

Turning attention to the role of the other SM elementary particles, we observe that in \req{FreeT} the finite $T$ contribution  is  non-negative, and is negligible for $gh\ll T$.  Therefore only SM particles with mass on the order of the phase transition temperature scale or higher will contribute.  Further study shows that particles with $g\gg T/v_0$ contribute equally to the effective potential except in a neighborhood of $h=0$.

The properties described above imply that including  particles with  GeV mass-scale and above in the free energy density $F$ assures that our considerations remain valid down to the scale $T\simeq 1 \GeV$. Thus we include the Higgs, gauge bosons  $W ^\pm,\, Z^0$ and the top, bottom, charm quarks,  and the tau lepton. The lighter particles, the $u,d,s$ quarks, gluons,   $e, \mu$ leptons as well as the neutrinos and photons are viewed as being massless in our discussion.

These `massless' SM particles  contribute  $0.1\%$ to the change in the effective potential and can be ignored in the study of phase transition properties. However, these particles contribute to the physical properties of the early Universe, an important application area of our considerations. Considering a phase transition in  laboratory experiments the leptonic degrees of freedom should be omitted as they  cannot be excited, only strongly coupled particles, i.e. quarks, gluons, and massive $Z^0, W^\pm$  are relevant.  

We approximate the effect of the Higgs by adding the free energy of a Bose gas to the effective potential in both the massless and massive phase, where the mass in each phase is determined by the Higgs effective potential due to all other particles.  This neglects the Higgs'  impact on self consistently determining its mass and the value of $h$ in each phase.  However, it is only one degree of freedom out of many so this effect should be negligible. 
 
\vskip 0.2cm 
%
{\bf Phase structure:}
Considering the properties of $U$, \req{TdepW}, we identify three critical temperatures relevant to the phase transition. We define\\
$\bullet$   $T_{c_1}$ to be the temperature where the massless phase is restored,\\
$\bullet$   $T_{c_2}$ to be the temperature at which the critical points become degenerate, and\\
$\bullet$   $T_{c_3}$ the temperature at which the massive phase is eliminated.\\
In the following, we will call $T_{c_2}$ the phase transition temperature. Note that by integrating \eqref{TdepW} by parts, one can see that the effective potential difference equals negative the pressure difference.  Therefore, for $T<T_{c_2}$ the massive phase is at a higher pressure than the massless phase, which implies that nucleated bubbles of the massive phase will expand.  The opposite is true for $T>T_{c_2}$ where the massless phase is at higher pressure and bubbles of symmetry restored vacuum expand.

The critical points occur at the zeros of 
\begin{align}
G(h,T)=&\frac{\lambda_hh(h^2-v_0^2)(1-B(k-1)(h^2/v_0^2-1)^2)}{(1+B(h^2/v_0^2-1)^2)^{k+1}}\notag\\[0.2cm]
&+\sum_j\frac{g_sg^2T^3h}{2\pi^2}\!\int_0^\infty\!\!\!\!\frac{z^2/E}{(e^{E/T}\pm 1)}\,dz
\end{align}
To track the location of the critical point for $T>0$ we solve the ode
\begin{equation}
\frac{dh}{dT}=-\frac{\partial G}{\partial T}\left(\frac{\partial G}{\partial h}\right)^{-1}.
\end{equation}

\begin{figure}
\includegraphics[height=9cm]{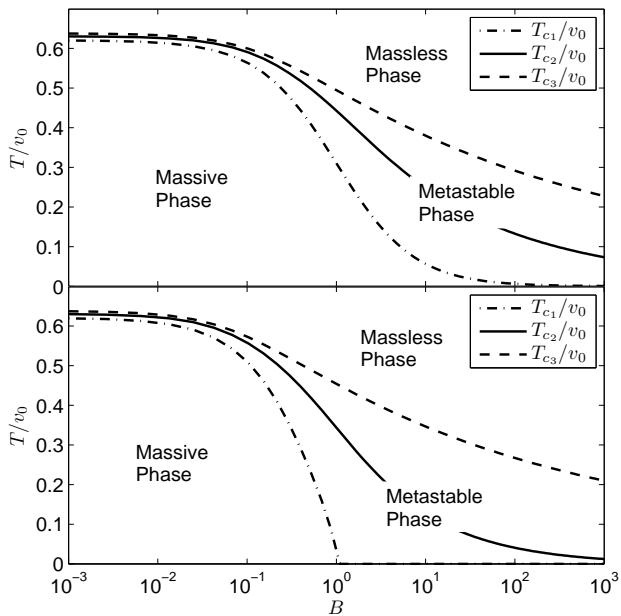}
\caption{Phase domains as a function of the Higgs potential parameter $B$  for $k=1$ (top) and $k=2$ (bottom).\label{fig:phase_diagram}}
\end{figure}

Figure~\ref{fig:phase_diagram} shows phase diagrams for $k=1$ (top) and $k=2$ (bottom) respectively. The solid line is where the two minima of the effective potential $U$, \req{TdepW}, are equal.  The dashed lines, as defined above, indicate where these minima first appear or respectively disappear depending on if we heat (heavy ion collisions) or cool (early Universe). We note that for $B\to 0$   both cases approach the limit of the SM Higgs potential, all three lines are very near to each other, converging at $T\simeq 0.63v_0$, corresponding to transformation temperature near 110 GeV. The phase transition at this level of discussion is weakly first order in agreement with  previous works, and quantum fluctuations  wash out this transition.

For either given value of $k$, the phase transition temperature, $T_{c_2}$, decreases as $B$ increases due to the reduction in $W(0)$. The large metastable region suggests that the phase transition is strongly first order for all $B>0$.  Also note that for $k=2$ and $B>1$ we are always in either the metastable or massless phase, even at zero temperature.  This holds under the more general condition, \req{Tc1_curve}.

\begin{figure}
\includegraphics[height=7cm]{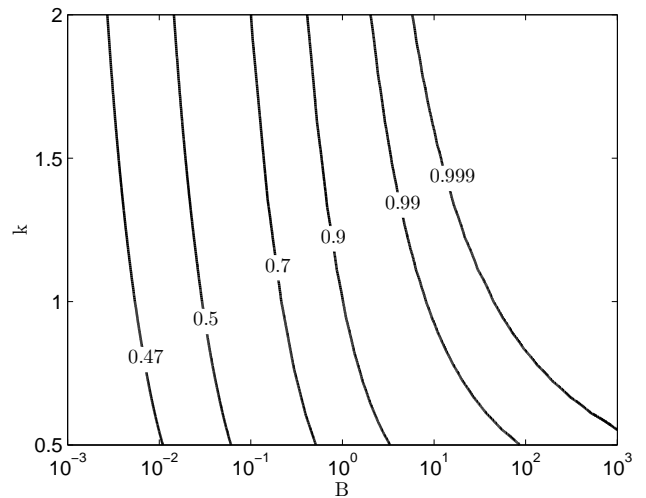}
\caption{$\Delta h/v_0$ at  critical temperature $T_{c_2}$ in the $k, B$ plane.\label{fig:transitionStrength}}
\end{figure}

\vskip 0.2cm 
%
{\bf Strength of the EW phase transition:}
As discussed above and seen in Fig. ~\ref{fig:phase_diagram} the transition is first order in the explored parameter range with coexistence domain increasing as $B$ increases. The transition is strong, in the sense that a large jump in Higgs and particle energy densities occurs when $h_{c_2}\approx v$ and $gh_{c_2}\gg T_{c_2}$, where $h_{c_2}$ denotes the value of $h$ in the massive phase at $T_{c_2}$.  Neither of these conditions hold for the standard Higgs potential, but as $B$ and $k$ increase, the strength of the phase transition increases.   Figure~\ref{fig:transitionStrength} shows the jump in $h/v_0$, which approaches unity for large $B>100$ considering value $k=1$. In such a transition there is a very large  change in the structure of the vacuum.

\begin{figure}
\includegraphics[height=8cm]{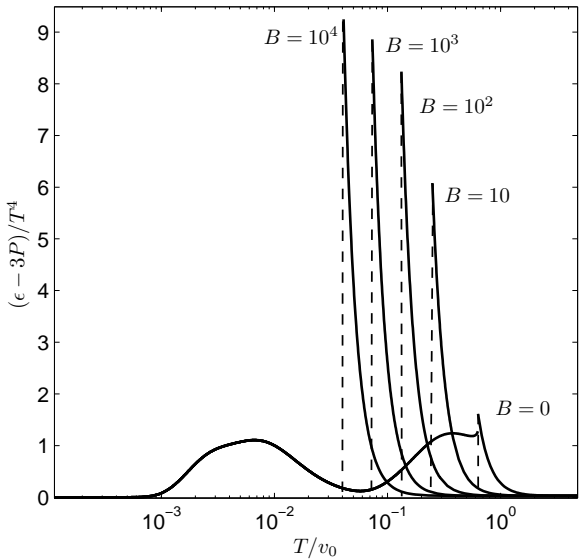}
\caption{The trace of the energy-momentum Tensor $\epsilon-3P$ scaled with $T^4$ as function of $T/v_0$ at $k=1$, and $B=0, 10, 10^2, 10^3, 10^4$.\label{fig:epsilon3Pk1.eps}}
\end{figure}

An illustration of the transition strength of relevance to the expandion of the Universe is the trace of the energy momentum tensor, $\epsilon-3P$, which is shown in Fig. \ref{fig:epsilon3Pk1.eps} scaled with $T^4$ as function of $T/v_0$.  The discontinuity is a consequence of the phase transition at $T_{c_2}$  For each value of $B$, the plot to the left and right of the discontinuity correponds to the massive and massless phases respectively.  For $B=0$ the minimum between the two humped structure originates in the temperature domain where all particles with mass below or at the bottom quark mass are nearly massless but the heavier $T, W^\pm, Z^0, H$ are not yet excited.

For the SM Higgs potential, $B=0$, we see in Fig.  \ref{fig:epsilon3Pk1.eps} that the SM Higgs case has a small discontinuity which is washed out by quantum fluctuation effects. In this case the expansion of the Universe seems smooth even though the expansion rate changes as it crosses the bumps in $(\epsilon-3P)/T^4$.  The positive jump in $\epsilon-3P$ as we cross the discontinuity is due to the contribution of the effective Higgs potential. This jump increases as $B$ does due to the lower transition temperature.  The energy density and pressure of the Higgs potential is independant of temeprature, and therefore dominates the contributions from the particles at lower $T$.

For the case of the strong transition $B>1$ a more complex dynamical picture ensues with a true first order phase transition appearing and a much more bumpy ride in the expansion, accompanied by nucleation dynamics which would require a significant effort to characterize. 
An important feature we see again in Fig. \ref{fig:epsilon3Pk1.eps} is that as $B$ increases the strong phase transition will occur at decreasing temperature. There is a priory no limit how far down we can push the temperature. Though the factor $B$ would need to be rather large to allow a transition in heavy ion experiments at the LHC, that is for $T/v_0=2\times10^{-3}$, there is no a priory reason to exclude this case. Recall here the relation of the two foundational scales,  $M_P/v_0=7\times 10^{16}$, where $M_P$ is the Planck Mass. Connections between these disparate scales have been explored and if there were to be a relation\cite{Bezrukov:2007ep}, appearance of a seemingly gigantic $B$ may not be that surprising.

\vskip 0.2cm 
%
{\bf Discussion:}
We have presented a class of Higgs effective potential that significantly lowers the EW phase transition temperature, see Fig.~\ref{fig:phase_diagram}, while preserving  the Higgs vacuum expectation value and mass. We further have shown that in a wide domain of parameters we considered, the transition is strongly first order, see Fig.~\ref{fig:transitionStrength},  and the transition temperature can be pushed to very low values by making an appropriate choice of the effective potential parameters.

The proposed modification of the (effective) Higgs potential seen in Fig.~\ref{fig:W_potential} has a vanishing impact on the properties of the Higgs particle in the massive phase, up to  vacuum fluctation effects. These do probe to some degree the structure of the potential beyond the location $\langle h\rangle=v_0$, though given the high Higgs mass these effects are small. However, in principle our proposed modification of the effective potential which comes along with a higher dimensional nonrenormalizable operator could influence precision fit of the properties of EW sector of the SM model. It would be of interest to see if the modification we propose are capable of improving the precision fit which tends to favor a Higgs mass below the experimentally favored value\cite{Nakamura:2010zzi}. Moreover, the range of effective potential parameters we consider could be perhaps constrained.

A vast domain of application of the proposed class of models is in cosmology.  To further the possibility of baryogenesis a related model was proposed earlier which assures a strong 1st order phase transition\cite{Grojean:2004xa}, and another approach with approximate conformal symmetry explored the effects of potential modifications further\cite{Konstandin:2011dr} . We have shown here by example that a 1st order EW transition can occur at a temperature scale 100 times smaller than previously considered possible, and thus relatively late in the evolution of the primordial Universe. The time span of potential baryogenesis\cite{Shaposhnikov:1994vm} associated with the EW phase transition is therefore considerably extended in view of the slow down of the expansion, potentially allowing for models consistent with observation. Interestingly, the  energy threshold for the breaking of baryon conservation can be expected to be much lower at the phase transiton temperatures  found here.  

We see potential opportunity to explore the electroweak phase transition in the laboratory. In the range $0.5<k<2$ and  for relatively large values of $B>100$  the transition temperature, barrier height, and latent heat are in the GeV range achievable in relativistic heavy ion collisions. A visible, natural signature of the formation of a massless EW phase in laboratory experiments would be abundant production of top and other massive quarks.

\vskip 0.2cm 
{\bf Acknowledgments:}
We thank M. Shaposhnikov for intersting discussions and valuable comments.
This work has been supported by the Department of Defense (DoD) through the National Defense Science \& Engineering Graduate Fellowship (NDSEG) Program and by a grant from the U.S. Department of Energy, DE-FG02-04ER41318 . 

%

\end{document}